# On some Implications of (Symmetric) Special Relativity to High Energy Physics

Ernst Karl Kunst
Im Spicher Garten 5
53639 Königswinter
Germany

**Beside the rise of total cross sections or interaction radii of colliding high energetic particles and the shrinkage of mean-free-paths of ultra relativistic particles (nucleii) in material media (anomalons), which have been shown to be of special relativistic origin [1], still other phenomena in high energy physics may arise from relativistic kinematics. In particular this seems to be the case with the EMC-effect and the so called atmosperic neutrino anomaly.**

In the mentioned work on relativistic kinematics has been shown a preferred rest frame of nature ($\Sigma_0$) in any inertial motion to exist and any velocity ($v_0$) be symmetrically composite or quantized. From this a symmetric modification of the Lorentz transformation follows between a frame of reference $S_1$ considered to be at rest according to the principle of relativity and a moving frame $S_2$

$$x_2' = \gamma_0(x_1 - v_0 t_1), \qquad y_2' = y_1, \qquad z_2' = z_1, \qquad t_2' = \gamma_0(t_1 - \frac{v_0}{c^2}x_1),$$

$$x_1^\circ = \gamma_0(x_2' + v_0 t_2'), \qquad y_1^\circ = y_2', \qquad z_1^\circ = z_2', \qquad t_1^\circ = \gamma_0(t_2' + \frac{v_0}{c^2}x_2'),$$

where

$$\gamma_0 = \left(1 - \frac{v_0^2}{c^2}\right)^{-\frac{1}{2}}.$$

The dashed symbols designate the moving system $S_2$ and the open circles the system $S_1$ at rest, now considered moving relative to $\Sigma_0$ and $S_2'$. Likewise the observer resting in $S_2$ will deduce the respective transformation:

$$x_1' = \gamma_0(x_2 + v_0 t_2), \qquad y_1' = y_2, \qquad z_1' = z_2, \qquad t_1' = \gamma_0(t_2 + \frac{v_0}{c^2}x_2),$$

$$x_2^\circ = \gamma_0(x_1' - v_0 t_1'), \qquad y_2^\circ = y_1', \qquad z_2^\circ = z_1', \qquad t_2^\circ = \gamma_0(t_1' - \frac{v_0}{c^2}x_1').$$

Furthermore, due to the absolute symmetry relative to $\Sigma_0$ must be valid:

$$x_2 = x_1, \quad y_2 = y_1, \quad z_2 = z_1, \quad t_2 = t_1,$$
$$x_1' = x_2', \quad y_1' = y_2', \quad z_1' = z_2', \quad t_1' = t_2',$$
$$x_2^\circ = x_1^\circ, \quad y_2^\circ = y_1^\circ, \quad z_2^\circ = z_1^\circ, \quad t_2^\circ = t_1^\circ$$

and always $|v_0| = |-v_0|$. If the upper lines of the of the above tansformation equations are inserted into the second lines, the identity results:

$$x_1^\circ \equiv x_1, \quad t_1^\circ \equiv t_1,$$
$$x_2^\circ \equiv x_2, \quad t_2^\circ \equiv t_2.$$

Further main results of the modified theory of relativistic kinematics among others are the Lorentz transformation not to predict the Fitzgerald-Lorentz contraction of the dimension (Δx) parallel to the velocity vector, as invented by Fitzgerald and Lorentz to account for the null-result of the Michelson-Morley experiment on moving Earth, but rather an expansion $\Delta x' = \Delta x \gamma_0$ - analogously to the relativistic time dilation $\Delta t' = \Delta t \gamma_0$. Accordingly the volume **V'** of an inertially moving body will any observer resting in a frame considered at rest seem enhanced

$$V' = \Delta x' \Delta y' \Delta z' = \Delta x \gamma_0 \Delta y \Delta z = V \gamma_0,$$

where **V** means volume. Among others it has been demonstrated, this expansion of Δx (or **V**) be the cause of the experimentally observed increase of the interaction radius respectively cross section of elementary particles with rising energy (velocity), as determined in collision experiments and as is known from studies of cosmic radiation, according to the equations

$$\overline{\sigma}_2' = \pi(\Delta r_1 \gamma_0^{\frac{1}{9}})^2 = \sigma_1 \gamma_0^{\frac{2}{9}}, \quad \overline{\sigma}_1' = \pi(\Delta r_2 \gamma_0^{\frac{1}{9}})^2 = \sigma_2 \gamma_0^{\frac{2}{9}} \tag{1}$$

so that the mean total geometrical cross-section is given by

$$\overline{\sigma}_{geo} = \overline{\sigma}_2' + \overline{\sigma}_1' = 2\overline{\sigma}_1 \gamma_0^{\frac{2}{9}}, \tag{2}$$

where $\bar{\sigma}_2 = \bar{\sigma}_1$, $\bar{\sigma}'_2 = \bar{\sigma}'_1$.
Hence the mean geometrical interaction- radius is given by

$$\overline{\Delta r'_2} = \Delta r_1 \gamma_0^{\frac{1}{9}}, \qquad \overline{\Delta r'_1} = \Delta r_2 \gamma_0^{\frac{1}{9}}. \qquad (3)$$

Because $v_0 \neq v$ (conventional velocity) it follows $\gamma_0 \neq \gamma$ (conventional Lorentz factor) so that predictions on the grounds of symmetric special relativity will deviate from the conventional view, the more the higher the velocity (see [1]).

### The Relativistic Origin of the "EMC-Effect"

The enhancement of the geometrical cross section or interaction radius according to the above equations also delivers an explanation of the so called EMC-effect in a direct way.
Consider the simplest case if the dimensionless variable $x \to 0$ so that the high energetic incident particle (electron, muon etc.) more or less traverses the nucleus, encountering on an average

$$\sqrt[3]{n_b}$$

nucleons within the nucleus. In this case the loss of momentum or energy of the outbound particle must be a relative one depending on the mean cross section the number $(n_b)^{1/3}$ of nucleons constituting the nucleus presents to the moving particle. Therefore, if ultra relativistic velocity or momentum per incident particle relative to the respective nucleus is assumed to be equal must the ratio of the relative dampening of momentum of the outwards moving particles within different nuclei independently of the respective ultra relativistc velocity or energy according to (2) be given by

$$R_{(x \to 0)} = \left( \sqrt[3]{\frac{n_b}{n'_b}} \right)^{\frac{2}{9}} = \left( \frac{n_b}{n'_b} \right)^{\frac{2}{27}},$$

whereby $n'_b > n_b$. Our formula delivers at $x = 0$ the ratios D/He = 0.95, D/C = 0.88, D/Al = 0.82, D/Ca = 0.80, C/Su = 0.84, which results agree very well with experiment [2],[3].
On the other end of the scale, where $x \to 1$, according to (3) the "shadowing effect" of the growing relative interaction radius of the respective nucleus relative to the incident particle must be considered. The scattering probability of the particle is dependent on the growing of the interaction radius of the respective target nucleus and, therewith, its dampening of momentum in dependence of the number of $(n_b)^{1/3}$ nucleons constituting the relative diameter of the nucleus at a ratio of

$$R_{(x \to 1)} = \left(\sqrt[3]{\frac{n_b}{n_b'}}\right)^{\frac{1}{9}} = \left(\frac{n_b}{n_b'}\right)^{\frac{1}{27}}.$$

Comparison with experiment also shows excellent correspondence [2],[3]. Thus, the EMC-effect is of the same physical (relativistic) origin as the rise of the total cross section or interaction radius of hadrons in high energetic collisions (see [1]).

### Relativistic Aberration (Doppler Boosting) as a Possible Cause of the Atmospheric Neutrino Results from Super-Kamiokande and Kamiokande

Consider a light signal or an ultra relativistic particle moving relative to $S'_2$ according to the equations

$$x'_2 = u_{x'_2} t'_2, \quad y'_2 = u_{y'_2} t'_2, \quad z'_2 = 0. \tag{4}$$

Transformation into the coordinates and the time of the moving system $S°_1$ delivers

$$u°_{x_1} = \frac{u_{x'_2} + v_0}{1 + \frac{v_0 u_{x'_2}}{c^2}}, \quad u°_{y_1} = \frac{u_{y'_2} \gamma_0^{-1}}{1 + \frac{v_0 u_{x'_2}}{c^2}}, \quad u°_{z_1} = 0,$$

wherefrom in connection with the above identity equations the aberration law of special relativity is deduced

$$\tan \alpha_1 = \frac{\sin \alpha'_2 \sqrt{1 - \frac{v_0^2}{c^2}}}{\cos \alpha'_2 + \frac{v_0}{c}}. \tag{5}$$

According to this theory (5) is valid as long as the systems $S°_1$ and $S'_2$ are considered freely moving relative to each other and no direct physical contact (collision) occurs. In the following will be shown that the relativistic aberration effect (5) predicts the short fall of muon neutrinos coming up through Earth, known as the atmosperic neutrino anomaly.
Experimental results from the Super-Kamiokande atmospheric neutrino measurements show at large distances from the neutrino generation, especially from Earth's far side, a significant suppression of the observed number of muon neutrinos with respect to the theoretical expectation [4]. For a relativistic analysis is of special interest that the muon neutrinos originate from two separate decay processes about 20 kilometers above: first a high energetic pion decays into a muon and a muon neutrino ($v_\mu$) and in a further step the muon into an electron, an electron neutrino ($v_e$) and a further muon neutrino. Thus, considering the decay modes only, the ratio of muon to electron neutrinos generated in the atmosphere can be predicted with confidence to be $R_{v(\mu/e)} = 2$. We underline especially the decay of the muon leading to the simultaneous generation of a $v_\mu$ and the $v_e$.

The Super-Kamiokande team compared particularly neutrinos coming down (downgoing) from the sky (l ≈ 20 km) with those coming upward (upgoing) through the Earth (l ≈ 12800 km). Because the cosmic rays and the resulting neutrinos rain down from all directions, the ratio should be $R = R_{v(\mu/e)upgoing}/R_{v(\mu/e)downgoing} = 1$. For electron neutrinos Super-Kamiokande caught equal numbers going up and coming down: $R_{v(e)upgoing}/R_{v(e)downgoing} \approx 1$, however, for muon neutrinos in 535 operation days 256 downward and only 139 upward ones have been counted. Furthermore, the expected ratio $R_{v(\mu/e)downgoing} \approx 2$ has been found, but only $R_{v(\mu/e)upgoing} \approx 1$, with systematic variations depending on the the distance l to the point of neutrino generation or angle of the incoming neutrinos. The number of muon neutrinos decreases linearly from a maximum at l ≈ 500 km to l ≈ 6400 - 7000 km, leveling off and remaining roughly constant up to l ≈ 12800 km at the far side of Earth.

This observed short fall of muon neutrinos with increasing distance l from the detector is currently interpreted as evidence for $v_\mu$ oscillations [4].

Consider a muon decaying at l ≈ 20 km above the detector. At the time of β-decay its ultra relativistic velocity $v_0$ may result in a Lorentz factor of $\gamma_0 = 10^4$ (which seems quite reasonable) and, to consider a simple case, the electron shall be emitted at some small angle « π/2 relative to the muon's direction. The neutrinos, counterbalancing the electron's momentum, will be emitted at larger angles $\alpha_{v(\mu)} > \alpha_{v(e)}$ if $p_{v(\mu)} > p_{v(e)}$ (which we assume). If for instance in the rest frame of the decaying muon for reasons of simplicity $\alpha_{v(e)} \approx 0°$ and $\Delta\alpha = \alpha_{v(\mu)} - \alpha_{v(e)} = 45°$ (90°, 135°), equation (5) predicts at l = 20 km a lateral displacement of 0.83 m (2 m, 4.83 m) and at l = 6400 km of 266 m (640 m, 1546 m) of the flight paths of both neutrino types. Even an improbably small angle of $\Delta\alpha = 5°$ (10°) in the rest frame of the muon would at l = 6400 km result in a lateral displacement of 28 m (60 m) between both neutrinos. A Lorentz factor of $10^5$ and an angle $\Delta\alpha = 45°$ (90°) would result in a lateral displacement of 0.08 m (0.2 m) at l = 20 km, 26.91 m (64 m) at l = 6400 km and 53 m (128 m) at l = 12800 km. If $\alpha_{v(e)} > 0°$ the respective differences have to be considered. It is clear that in a broad band of varying angles, muon energy or Lorentz factors we would for distances ≳ 6400 km arrive at similiar results:

$$\arctan\left[\gamma_0\left(\frac{\sin\alpha_{v_\mu}}{1+\cos\alpha_{v_\mu}} - \frac{\sin\alpha_{v_e}}{1+\cos\alpha_{v_e}}\right)\right] \times \frac{2\pi l}{360} \geq d,$$

where d means diameter of the detector, $\alpha_{v(\mu)}$ and $\alpha_{v(e)}$ are given in degrees and $v_0/c = 1$. Even in the case of a very high energy muon would a respective large angle Δα between the tracks of both neutrino types in its rest frame according to the above examples at l = 6400 km lead to a lateral displacement > d (Super-Kamiokande: d = 34 m and hight = 36 m).

It is clear that in all these cases at l ≈ 20 km the muon-type and electron-type neutrino would traverse jointly a sufficiently large detector so that for each of the two neutrino types there is an equal chance to become counted, whereas at l ≥ 6400 km for each counted electron neutrino the accompanying muon neutrino will due to the large displacement mainly miss the detector and pass far away undetectably by.The exact distance dependence of this relativistic aberration effect also explains the linear decrease of the number of muon neutrinos with increasing distance from the maximum of counts to the point, wherefrom most upgoing muon neutrinos no more can reach the detector together with the electron neutrino in l ≥ 6400 km, in a fully way.

Thus, if, as already mentioned, the flux of upgoing and downgoing electron neutrinos is observed to be about equal so that $R_{v(e)upgoing}/R_{v(e)downgoing} \approx 1$, this conclusively implies that from the far side of Earth mainly those muon neutrinos reach the detector, which were generated by the pion decay. All or nearly all muon neutrinos generated together with the electron neutrinos in the course of the muon decay on the other hand can due to the growing distance from the electron neutrino's flight path not be registered by the same detector, resulting in $R_{v(\mu/e)upgoing} \approx 1$, exactly as observed.

**Transversal Aberration Effects in Ultra Relativistic Collision Events**

The theory also explains independently of quantum mechanical models the steady increase of the mean transverse energy per particle and unexpected frequent appearance of events with very high transverse energy as well as their distribution normal to the beam direction in collision experiments. But owing to the strong relativistic elongation of the colliding particles in beam direction this transversal aberration effect will at lower energies be superimposed by a longitudinal alignment of the secondary particles - in agreement with experiment.

Consider two identical particles in $S'_2$ and $S''_2$ colliding elastically with equal but oppositely directed velocity at point $S_1$ at rest, being the kinematical center and at the same time the center-of-mass. We restrict our analysis to the recoil particle in $S'_2$. In the time particle dt after collision, $S_1$ is moving relative to $S'_2$ according to the equations (4). Transformation into the coordinates and time of $S_1$ delivers

$$u_{x_1} = \frac{u_{x'_2} + v_0}{1 + \frac{v_0 u_{x'_2}}{c^2}}, \quad u_{y_1} = \frac{u_{y'_2}/\gamma_0}{1 + \frac{v_0 u_{x'_2}}{c^2}}, \quad u_{z_1} = 0.$$

Thus, one expects an aberrational transversal deviation of ultra relativistic recoil particles and photons from the path pattern in the center-of-mass, which is given by

$$\tan\alpha_1 = \frac{\sin\alpha_2'}{\left(\cos\alpha_2' + \frac{v_0}{c}\right)\sqrt{1 - \frac{v_0^2}{c^2}}} \ . \tag{6}$$

This transversal deviation in collisions is a purely relativistic effect in the kinematical center, which only depends on the velocity of the colliding particles (of equal mass). Division of the tangens function at two different velocities delivers

$$R_T = \gamma_0' \gamma_0^{-1}, \tag{7}$$

where $\gamma_0' < \gamma_0$, implying $v_0' > v_0$. This ratio is of interest in extrapolating the increase of transversal momentum (energy) at different velocities and scattering angles in collision experiments in the center-of-mass frame (see below).

If the collision of an ultra relativistic particle with a fixed target particle in the laboratory is analysed, evidently neither system, the particles rest within, can be considered at rest owing to the natural rest frame $\Sigma_0$ amidst them, implying both systems to move relative to each other. Obviously this requirement is fulfilled if the laboratory is considered as the moving system $S_1°$ and the ultra relativistic particle as the oppositely moving system $S_2°$ according to the above transformation equations. According to the latter also is valid $x_2° = x_1°$ etc. so that follows

$$\tan\alpha_1° = \tan\alpha_2°. \tag{8}$$

Thus, no relativistic deviation is to expect in collision events with fixed targets. If we consider the above identity equations and transform the right hand member of (8) into the coordinates and time of $S_2'$, we receive again

$$\tan\alpha_1 = \frac{\sin\alpha_2'}{\left(\cos\alpha_2' + \frac{v_0}{c}\right)\sqrt{1 - \frac{v_0^2}{c^2}}},$$

the apparent increase of energetic particles transversal to the beam direction with growing velocity (energy), as compared with the expectations on the grounds of the validity of the aberration law (5) of special relativity for this kind of scattering experiment.
In the following predictions on the grounds of (7) are compared with experiment.

The differential cross section is defined by

$$\frac{d\sigma}{d\Omega} = \frac{\text{number of particles at scattering} \angle \theta, \varphi/\text{time} \times \text{solid} \angle}{\text{current density of incident particles} \times \text{number of scattering centers}}.$$

Irrespective of quantum mechanical effects the "number of particles at scattering angle θ" must directly depend on the rise of the geometrical cross section according to (1) and (2) and the transversal aberration effect in ultra relativistic collisions according to (6) in the case of colliders as well as accelerators. Therefore, at low transverse momentums the differential cross sections of scattering experiments at higher energies are extrapolatable with fair accuracy from low energy values. For this purpose simply the product of the ratio of the geometrical cross section at lower and higher velocity (energy), and of the the ratio (7) - the aberrational effects in collision events predicted by this theory as compared with special relativity - is to multiply by the tangens of the transverse momentum:

$$\tan \sigma_T = R_T \gamma_0'^{\frac{2}{9}} \gamma_0^{-\frac{2}{9}} \tan p_T, \tag{9}$$

where $p_T$ means a given transverse momentum at lower velocity ($p_T \leq 10$ GeV) and $\gamma_0' < \gamma_0$. This simple geometrical derivation of aberrational effects at lower momentums from experimental values is possible because the differential cross section at a given velocity (energy) as a function of the four-momentum transfer squared is equivalent to plotting it as a function of scattering angle at fixed energy. The curve of the differential cross section at higher energies is geometrically approximated by the above formula by adding arctan $\sigma_T$ units to the point $p_T$ of the curve at lower scattering velocity (energy). The symmetrical transverse momentum $p_T$ is computed from the conventional momentum (see [1]). In the figure experimental results at the center-of-mass energy E* = 540 GeV are compared with extrapolations(crosses) from E* = 62 GeV to E* = 540 GeV according to the above formula. The approximation seems fairly good.

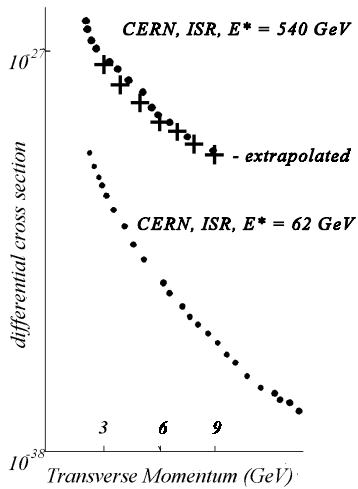

Thus, it is predicted that the growing transversal deviation of the secondary particles out of collisions with ever growing energy of particles of whatever kind, as for instance found at the Fermilab's Tevatron particle accelerator in protron-antiprotron collisions at 1800 GeV in the center-of-mass frame, is solely of relativistic origin. This also is true for high energetic collisions of all kinds of nucleii, where indeed this trend has been observed long since, as for instance at GSI in Darmstadt, Germany. In violent collisions (1 GeV/nucleon) between gold nuclei has been found that at polar emission angles of 90° in the center-of-mass frame, kaons as well as nucleons and pions emerge preferentially out of the plane of the collision, although kaons are expected to emerge isotropically [5]. However, according to the above formulas this effect clearly is to expect and must increase with ever growing energy (velocity).

According to this theory the experimentally verified tendency of secondaries in high energetic collisions to deviate transversally to the beam direction with increasing energy (velocity) is merely of relativistic origin. This effect also comprises "jet" structures "seen" in protron-protron (antiproton) and electron-positron collisions, which usually are interpreted as a manifestation of the interaction of the quarks constituting the hadrons. But according to this theory the observed jet structures are (mainly) fictitious and necessarily occur if $v_0 \to c$ and the emitted particles - independently of their origin (elastic or inelastic scattering) - according to the above equations tend to fill out the transversal region. If high transverse momentums and jets are of the same kinematic origin, they also should exhibit similiar structures, regardless of the particles involved. Furthermore is clear that the bulk of high transverse energy events should have a non-jet like uniform azimuthal distribution. Indeed this is observed. Therefore, the probability of the production rate of jets should rise in accordance with this theory independently of the particles involved. And indeed: the UA1 experiment at CERN observed a rise of the jet cross section in protron-antiprotron collisions from $\approx$ 5 mb at 350 GeV (center-of-mass frame) collision energy to $\approx$ 10 mb at 900 GeV [6]. Extrapolation according to (9) also results in 10 mb.

Comparable data were measured by the PLUTO experiment at DESY in Hamburg in electron-positron collisions. A rise of the mean square sums $(p_T)^2$ of the transversal momentum of jets as a function of the center-of-mass energy E from $\approx$ 1.3 $<\Sigma(p_T)^2>(GeV^2)$ at E = 7.7 GeV to $\approx$ 6.6 $<\Sigma(p_T)^2>(GeV^2)$ at E = 31.6 GeV has been observed [7]. Extrapolation according to (9) results in 6.6 $<\Sigma(pT)^2>(GeV^2)$, too.

It is clear that the mean total transverse energy (the sum of all $E_T$s in an event) or mean transversal momentum must rise proportionally to the relativistic rise of the mean geometrical cross section in connection with the transversal aberration. Respective measurements were made at CERN, where a rise of the mean total transverse energy from $\approx$ 300 MeV at ISR energy (60 GeV) to $\approx$ 500 MeV at SPS energy (540 GeV) has been observed [8]. Extrapolation results in

$$300 \; MeV \times \left( \frac{E'_{(E^*=540 \; GeV)} \left( E'_{(E^*=540 \; GeV)} \right)^{\frac{2}{9}}}{E'_{(E^*=60 \; GeV)} \left( E'_{(E^*=60 \; GeV)} \right)^{\frac{2}{9}}} \right)^{\frac{2}{9}} = 456 \; MeV.$$

At CERN in classic protron-antiprotron scattering events a rise of the mean cross section of the individual particles transversal to the beam direction from 56.1 (±4.7) mb at 150 GeV total energy in the center-of-mass system to 85.5 (±6.4) mb at 900 GeV has been measured [9]. If the cross section of the protron (antiprotron) at rest = 10 mb our formulas deliver:

$$\sigma_{T(150\ GeV)} \approx 20\ mb \times \left( E'_{(E^*=150\ GeV)} \left( E'_{(E^*=150\ GeV)} \right)^{\frac{2}{9}} \right)^{\frac{2}{9}} = 59.28\ mb,$$

$$\sigma_{T(900\ GeV)} \approx 20\ mb \times \left( E'_{(E^*=900\ GeV)} \left( E'_{(E^*=900\ GeV)} \right)^{\frac{2}{9}} \right)^{\frac{2}{9}} = 84.83\ mb,$$

where T in the left hand side means transversal, E' energy based on (quantized) $v_0$ and E* center-of-mass energy (see [1]).
Respective measurements of electron-positron collisions were carried out by the PLUTO experiment at DESY in Hamburg. A rise of the mean transversal momentum of jets from ≈ 0.3 GeV at 7.7 GeV total energy in the center-of-mass system to ≈ 0.4 GeV at 31.6 GeV collision energy has been observed [10]. Extrapolation also results in 0.4 GeV and, thus, shows very good correspondence with experiment.

Finally it is predicted that the excess of events (as compared with the expectations on the grounds of the standard model) in collisions between positrons of 27.5 GeV and protons of 820 GeV center-of-mass energy with high momentum transfer or at a large angle, found at DESY's HERA positron-proton collider, also owes its existence to the relativistic rise of the mean geometrical cross section in connection with relativistic transversal aberration in collisions.